\documentclass[onecolumn,showpacs]{revtex4}

\usepackage{graphicx}
\usepackage{dcolumn}
\usepackage{bm}

\input epsf

\begin{document}

\title{\Large Gravitational Collapse in Higher Dimension}

\author{\bf Ujjal Debnath}
\email{ujjaldebnath@yahoo.com}
\author{\bf Subenoy Chakraborty}
\email{subenoyc@yahoo.co.in}
\affiliation{Department of
Mathematics, Jadavpur University, Calcutta-32, India.}

\date{\today}

\begin{abstract}
Spherically symmetric inhomogeneous dust collapse has been
studied in higher dimensional space-time and appearance of naked
singularity has been analyzed both for non-marginal and
marginally bound cases. It has been shown that naked singularity
is possible for any arbitrary dimension in non-marginally bound
case. For marginally bound case we have examined the radial null
geodesics from the singularity and found that naked singularity
is possible upto five dimension.
\end{abstract}


\maketitle

\section{\normalsize\bf{Introduction}}
For the last two decades or so, gravitational collapse is an
important and challenging issue in Einstein gravity, particularly
after the formation of famous singularity theorems [1] and Cosmic
Censorship Conjecture (CCC) [2]. Also it is interesting to know
the final outcome of gravitational collapse [3] in the background
of general relativity from the perspective of black hole physics
as well as its astrophysical implications. The singularity
theorems provide us only about the generic property of space
times in classical general relativity but these theorems can not
predict the detailed features of the singularities such as their
visibility to an external observer as well as their strength. The
CCC, on the other hand is incomplete [4,5] as it stands because
there is no formal proof of it in one hand and on the other there
are counter examples of it. However, the nature of the central
shell focusing singularity depends on the choice of the initial
data [6].\\

It has recently been pointed out by Joshi et al [7] that the
physical feature which is responsible for the formation of naked
singularity is nothing but the presence of shear. It is the shear
developing in the gravitational collapse, which delays the
formation of the apparent horizon so that the communication is
possible from the very strong gravity region to observers
situated outside.\\

The objective of this paper is to fully investigate the situation
in the background of higher dimensional space-time with both
non-marginally and marginally bound collapse [8-10]. As it is only
to be expected, in one way or another, these works all deal with
propagation of null geodesics in the space-time of a collapsing
dust [12]. In this context we mention that Ghosh and Beesham [10]
have also studied dust collapse for (n+2) dimensional
Tolman-Bondi space-time for marginally bound case ($f=0$),
considering the self-similar solutions. They have concluded that
higher dimensions are favourable for black holes rather than
naked singularities. Also recently, Ghosh and Banerjee [11] have
considered non-marginal case ($f\ne 0$) for dust collapse in 5D
Tolman-Bondi model and have shown that the degree of
inhomogeneity of the collapsing matter is necessary to form a naked singularity. \\

We show in the present paper that in non-marginally bound collapse
(i.e., for $f\ne 0$), the naked singularity may appear in any
dimensional space-time, but for marginally bound collapse the
naked singularity may appear only when the space-time has
dimensions upto five.\\

\section{\normalsize\bf{Non-marginally bound case}}
The metric ansatz for $n$ dimensional space time is

\begin{equation}
ds^{2}=-dt^{2}+\frac{R'^{2}}{1+f(r)}dr^{2}+R^{2}d\Omega^{2}_{n-2}
\end{equation}

with~~
$d\Omega^{2}_{n-2}=d\theta^{2}_{1}+Sin^{2}\theta_{1}[d\theta^{2}_{2}+Sin^{2}\theta_{2}
(d\theta^{2}_{3}+......+Sin^{2}\theta_{n-3}d\theta^{2}_{n-2})]$\\

We consider here a dust collapse with energy-momentum tensor:

\begin{equation}
T_{\mu\nu}=\rho(t,r) u_{\mu} u_{\nu} ,
\end{equation}

with $u_{\mu}$ as the $n$-velocity.\\

Now from the Einstein's field equations for the metric (1) with
the above energy-momentum tensor, one can obtain (choosing $8\pi
G=c=1$)

\begin{equation}
\dot{R}^{2}=\frac{F(r)}{R^{n-3}}+f(r),
\end{equation}

and

\begin{equation}
\rho(t,r)=\frac{(n-2)F'(r)}{2R^{n-2}R'},
\end{equation}

where $F(r)$ is an arbitrary functions of $r$, arising from the
integration with respect to the proper time $t$.\\

Suppose, the collapse develops from an initial surface $t=t_{i}$
and the above model is characterized by the initial density
$\rho_{i}(r)=\rho(t_{i},r)$ and $f(r)$, which describes the
initial velocities of collapsing matter shells. We choose the
scaling of the scale factor $R$ such that

\begin{equation}
R(t_{i},r)=r,
\end{equation}

so that

\begin{equation}
\rho_{i}(r)=\rho(t_{i},r)=\frac{n-2}{2}r^{2-n}F'(r)
\end{equation}

Now the curve $t=t_{s}(r)$ defines the shell-focusing singularity
and is characterized by

\begin{equation}
R(t_{s}(r),r)=0
\end{equation}

Further, within the collapsing cloud, the trapped surfaces will
be formed due to unbounded growth of the density and these
trapped surfaces are characterized by the outgoing null
geodesics. In fact, the apparent horizon, which is the boundary
of the trapped surface has the equation $t=t_{ah}(r)$ and the
scale factor at the apparent horizon satisfies

\begin{equation}
R(t_{ah}(r),r)=[F(r)]^{\frac{1}{n-3}}
\end{equation}

To characterize the nature of the singularity we shall discuss
the two possibilities namely, $(i)$ $t_{ah}<t_{s}(0)$ and $(ii)$
$t_{ah}>t_{s}(0)$. The first case may correspond to formation of
black hole while the second one may lead to naked singularity. If
the apparent horizon will form earlier than the instant of the
formation singularity, then the event horizon can fully cover the
strong gravity region and also the singularity. As a result, no
light signal from the singularity can reach to any outside
observer and the singularity is totally hidden within a black
hole. On the other hand, in the second case the trapped surfaces
will form much later during the evolution of the collapse and it
is possible to have a communication between the singularity and
external observers.\\

Integrating once, we have from equation (3)

\begin{equation}
t-t_{i}=\frac{2}{(n-1)\sqrt{F}}\left[r^{\frac{n-1}{2}}~_{2}F_{1}[\frac{1}{2},a,a+1,-\frac{f
r^{n-3}}{F}]-R^{\frac{n-1}{2}}~_{2}F_{1}[\frac{1}{2},a,a+1,-\frac{f
R^{n-3} }{F}]\right]
\end{equation}

where we have used the initial condition (5) and $_{2}F_{1}$ is
the usual hypergeometric function with
$a=\frac{1}{2}+\frac{1}{n-3}$. Using equations (7) and (8)
separately in equation (9) we have

\begin{equation}
t_{s}(r)-t_{i}=\frac{2}{(n-1)\sqrt{F}}r^{\frac{n-1}{2}}~_{2}F_{1}[\frac{1}{2},a,a+1,-\frac{f
r^{n-3}}{F}]
\end{equation}

and

\begin{equation}
t_{ah}(r)-t_{i}=\frac{2r^{\frac{n-1}{2}}}{(n-1)\sqrt{F}}~_{2}F_{1}[\frac{1}{2},a,a+1,-\frac{f
r^{n-3}}{F}]-\frac{2F^{\frac{1}{n-3}}}{n-1}~_{2}F_{1}[\frac{1}{2},a,a+1,-f]
\end{equation}

We note that in case of homogeneous dust, the collapse is
simultaneous (Openheimer-Snyder) but in the present inhomogeneous
model the collapse is not simultaneous (in comoving co-ordinates)
but rather the singularity is described by a curve with starting
point $(t_{0},0)$, which is given by equation (10) as
\begin{eqnarray}
t_{0}=t_{s}(0)=t_{i}+
\begin{array}{ll}lim~~~~\frac{2r^{\frac{n-1}{2}}}{(n-1)\sqrt{F}}~_{2}F_{1}[\frac{1}{2},a,a+1,-\frac{f
r^{n-3}}{F}]\\
r\rightarrow 0
\end{array}
\end{eqnarray}

Now in order to have a finite value of the above limit, we assume
$F(r)$ and $f(r)$ [13,14] to be in the following polynomial form
near the central singularity $(r=0)$

\begin{equation}
F(r)=F_{0}r^{n-1}+F_{1}r^{n}+F_{2}r^{n+1}+.........
\end{equation}
and
\begin{equation}
f(r)=f_{0}r^{2}+f_{1}r^{3}+f_{2}r^{4}+.........
\end{equation}

Then equation (5) suggests that initial density profile is also
smooth at the centre and we have

\begin{equation}
\rho_{i}(r)=\rho_{0}+\rho_{1}r+\rho_{2}r^{2}+.........
\end{equation}

with ~~$\rho_{j}=\frac{(n+j-1)(n-2)}{2}F_{j},~~~j=0,1,2,...$~.\\

As the density gradient is negative and falls off rapidly to zero
near the centre so we must have $\rho_{1}=0$ and $\rho_{2}<0$ and
consequently $F_{1}=0$ and $F_{2}<0$. Thus the above limit (see
eq.(12)) simplifies to

\begin{equation}
t_{0}=t_{i}+\frac{2}{(n-1)\sqrt{F_{0}}}~_{2}F_{1}[\frac{1}{2},a,a+1,-\frac{f_{0}
}{F_{0}}]
\end{equation}

Therefore, equation (11) and (16) give (after using (13) and (14)
in (11))

\begin{eqnarray*}
t_{ah}-t_{0}=-\frac{2}{n-1}F_{0}^{\frac{1}{n-3}}r^{\frac{n-1}{n-3}}
-\frac{f_{1}}{(3n-7)F_{0}^{3/2}}~_{2}F_{1}[\frac{3}{2},a+1,a+2,-\frac{f_{0}
}{F_{0}}]r+\left[-\frac{F_{2}}{(n-1)F_{0}^{3/2}}\right.
\end{eqnarray*}
\vspace{-5mm}

\begin{eqnarray*}
\left._{2}F_{1}[\frac{1}{2},a,a+1,-\frac{f_{0}
}{F_{0}}]+\frac{1}{4(n-3)(3n-7)f_{0}F_{0}^{5/2}(f_{0}+F_{0})}\left\{(7-3n)f_{0}^{2}F_{0}^{2}
~_{2}F_{1}[\frac{1}{2},a,a+1,-\frac{f_{0} }{F_{0}}]\right.\right.
\end{eqnarray*}
\vspace{-5mm}

\begin{eqnarray*}
\left.\left.+\{
-2f_{0}F_{0}((5-2n)f_{1}^{2}+2(n-3)f_{2}F_{0})-4(n-3)f_{0}^{2}F_{0}(f_{2}-F_{2})
+4(n-3)f_{0}^{3}F_{2}\right.\right.
\end{eqnarray*}
\vspace{-8mm}

\begin{equation}
\left.\left.+(3n-7)f_{1}^{2}F_{0}^{2} \}
~_{2}F_{1}[\frac{3}{2},a+1,a+2,-\frac{f_{0} }{F_{0}}] \right\}
\right]r^{2}+O(r^{3})
\end{equation}

From the above relation we note that near $r=0$, $t_{ah}> or \le
0$ according as
$$f_{1}~_{2}F_{1}[\frac{3}{2},a+1,a+2,-\frac{f_{0}}{F_{0}}]< or \ge
0 .$$ Though the hypergeometric function has arguments depending
on $n$ yet $_{2}F_{1}$ is always positive. So the above
restriction simply implies $f_{1}< or \ge 0$. Now according to
Joshi etal, if a comoving observer (at fixed $r$) does not
encounter any trapped surfaces until the time of singularity
formation then it is possible to visualize the singularity
otherwise the singularity is covered by trapped surfaces, leading
to a black hole. Thus $f_{1}<0$ leads to naked singularity while
$f_{1}\ge 0$ is the restriction for black hole. However, the
above conditions are not sufficient, particularly for non-smooth
initial density profile. Therefore, we may have locally naked
singularity or black hole depending on the sign of $f_{1}$.\\

Further, to study the effect of shear on the formation of trapped
surface we first evaluate the shear explicitly. For
$n$-dimensional spherically symmetric dust metric the shear
scalar is estimated by [7]
$$
\sigma=\sqrt{\frac{n-2}{2(n-1)}}\left(\frac{\dot{R}'}{R'}-\frac{\dot{R}}{R}\right)
$$

Using equation(3) we get in turn

\begin{equation}
\sigma=\sqrt{\frac{n-2}{8(n-1)}}~\frac{\left[\{R F'-(n-1)R' F
\}+R^{n-3}(R f'-2R' f)\right]}{R^{\frac{n-1}{2}}R'\left(F+f
R^{n-3} \right)^{1/2}}
\end{equation}

Since at the initial hypersurface ($t=t_{i}$) we have chosen
$R(t_{i},r)=r$, so the initial shear $\sigma_{i}$ is of the form
$$
\sigma=\sqrt{\frac{n-2}{8(n-1)}}~\frac{\left[\{r F'-(n-1)F
\}+r^{n-3}(r f'-2f)\right]}{r^{\frac{n-1}{2}}\left(F+f r^{n-3}
\right)^{1/2}}
$$

Thus using the power series expansions (13) and (14) for $F(r)$
and $f(r)$ in the above expression for $\sigma_{i}$ we have

\begin{eqnarray*}
\sigma_{i}=\sqrt{\frac{n-2}{8(n-1)}}~\frac{f_{1}r+\sum_{m=2}^{\infty}m(f_{m}+F_{m})r^{m}}
{\sqrt{(f_{0}+F_{0})+f_{1}r+\sum_{m=2}^{\infty}m(f_{m}+F_{m})r^{m}}}
\end{eqnarray*}
\vspace{-5mm}

\begin{equation}
\hspace{1.18in}=\sqrt{\frac{n-2}{8(n-1)}}~\frac{1}{\sqrt{f_{0}+F_{0}}}~\left[f_{1}r+
\left\{2(f_{2}+F_{2})-\frac{f_{1}^{2}}{2(f_{0}+F_{0})}\right\}r^{2}+O(r^{3})\right]
\end{equation}

We note that the initial shear vanishes when $f_{1}=0,
(F_{2}+f_{2})=(F_{3}+f_{3})=......=0$ and hence even if the
initial shear is zero the dust distribution may be inhomogeneous.
Thus from equation (10) $t_{s}$ a function of the comoving radial
co-ordinate $r$, so that the shell focusing singularity appears
at different $r$ at different instants. Also the above expression
for $\sigma_{i}$ reveals that the existence of the naked
singularity is not directly related to the non-vanishing of the
shear as it does in the marginally bound case (see ref.[14]).\\

\section{\normalsize\bf{Marginally bound case}}
In this case (i.e., $f=0$) equation (3) can be integrated out
easily to give

\begin{equation}
R^{\frac{n-1}{2}}=r^{\frac{n-1}{2}}-\frac{n-1}{2}\sqrt{F(r)}(t-t_{i})
\end{equation}

where we have used the initial condition (5).\\

Suppose the radius of the spherical shell $R$ shrinks to zero at
the time $t_{c}(r)$ then from (20) we have

\begin{equation}
t_{c}(r)-t_{i}=\frac{2}{n-1}\frac{r^{\frac{n-1}{2}}}{\sqrt{F(r)}}
\end{equation}

Now the Kretchmann scalar

\begin{equation}
K=[(n-2)(n-3)+1]\frac{F'^{2}}{R^{2n-4}R'^{2}}-2(n-2)^{2}(n-3)\frac{FF'}{R^{2n-3}R'^{2}}+
(n-1)(n-2)^{2}(n-3)\frac{F^{2}}{R^{2n-2}}
\end{equation}

diverges at $t=t_{c}(r)$ i.e., $R=0$. Thus it represents the
formation of a curvature singularity at $r$. In fact the central
singularity (i.e., $r=0$) forms at the time

\begin{equation}
t_{0}=t_{i}+\sqrt{\frac{2(n-2)}{(n-1)\rho_{_{0}}}}
\end{equation}

The Kretchmann scalar also diverges at this central
singularity.\\

Now if we use the expansion (13) for $F(r)$ in equation (21) then
near $r=0$, the singularity curve can be approximately written as
(using (23))

\begin{equation}
t_{c}(r)=t_{0}-\frac{F_{m}}{(n-1)F_{0}^{3/2}}r^{m}
\end{equation}

where $m\ge 2$ and $F_{m}$ is the first non-vanishing term beyond
$F_{0}$. Thus $t_{c}(r)>t_{0}$ as $F_{m}<0$ for any $m\ge 2$.\\

To examine whether the singularity at $t=t_{0},r=0$ is naked or
not, we investigate whether there exist one or more outgoing null
geodesics which terminate in the past at the central singularity.
In particular, we will concentrate to radial null geodesics
only.\\

Let us start with the assumption that it is possible to have one
or more such geodesics and we choose the form of the geodesics
(near $r=0$) as

\begin{equation}
t=t_{0}+a r^{\alpha},
\end{equation}

to leading order in $t$-$r$ plane with $a>0,\alpha>0$. Now for
$t$ in the geodesic (25) should be less than $t_{c}(r)$ in (24)
for visibility of the naked singularity so on comparison we have

\begin{equation}
\alpha\ge m ~~~~ \text{and} ~~~~a<-\frac{F_{m}}{(n-1)F_{0}^{3/2}}.
\end{equation}

Also from the metric form (1), an outgoing null geodesic must
satisfy

\begin{equation}
\frac{dt}{dr}=R'
\end{equation}

But near $r=0$, the solution (20) for $R$ simplifies to

\begin{equation}
R=r\left[1-\frac{n-1}{2}\sqrt{F_{0}}\left(1+\frac{F_{m}}{2F_{0}}r^{m}
\right)t \right]^{\frac{2}{n-1}}
\end{equation}

Thus combining (25) and (28) in equation (27) we get

\begin{equation}
a \alpha r^{\alpha-1}=\frac{\left[1-\frac{n-1}{2}\sqrt{F_{0}}~
(t_{0}+a
r^{\alpha})-\frac{(2m+n-1)F_{m}}{4\sqrt{F_{0}}}r^{m}(t_{0}+a
r^{\alpha})
\right]}{\left[1-\frac{n-1}{2}\sqrt{F_{0}}\left(1+\frac{F_{m}}{2F_{0}}r^{m}
\right)(t_{0}+a r^{\alpha}) \right]^{\frac{n-3}{n-1}}}
\end{equation}

Now if there exists a self consistent solution of this equation
then it is possible to have at least one outgoing radial null
geodesic that had started at the singularity i.e., the
singularity is naked. In order to simplify the above equation we
shall use the restrictions in equation (26) in the following two
ways:\\

$(i)~~ \alpha>m$ :\\

The equation (29) becomes (in leading order)

\begin{equation}
a \alpha
r^{\alpha-1}=\left(1+\frac{2m}{n-1}\right)\left(-\frac{F_{m}}{2F_{0}}\right)
^{\frac{2}{n-1}}r^{\frac{2m}{n-1}}
\end{equation}

which implies

\begin{equation}
\alpha=1+\frac{2m}{n-1}~~~ \text{and}~~~
a=\left(-\frac{F_{m}}{2F_{0}}\right)^{2/(n-1)}
\end{equation}

Thus for $\alpha>m$ we have $m<\frac{n-1}{n-3}$ and $n>3$.\\

For $n=4$, $m$ may take values 1 and 2 and we have
$\rho_{1}=F_{1}=0$ and $\rho_{2}<0,F_{2}<0$. As a result, $a$ is
real and positive from equations (26) and (31). Moreover, these
restrictions are already assumed in the power series expansion
for $\rho_{i}(r)$ so that the initial density gradient is
negative and falls off rapidly near the centre. But for $n>4$,
$m=1$ is the only possible solution for which no real positive
solution of `$a$' is permissible from equations (26) and (31).
Hence with the restriction $\alpha>m$, we have a consistent
solution of equation (29) only for four dimensional space-time
i.e., it is possible to have (at least) null geodesics terminate
in the past at the singularity only for four dimension and we can
have naked singularity for $n=4$.\\

$(ii)~~\alpha=m$ :\\

In this case equation (29) simplifies to

\begin{equation}
m a
r^{m-1}=\left[-\frac{F_{m}}{2F_{0}}-\frac{n-1}{2}\sqrt{F_{0}}a\right]^{\frac{3-n}{n-1}}
\left[-\frac{(2m+n-1)}{2(n-1)}\frac{F_{m}}{F_{0}}-\frac{n-1}{2}\sqrt{F_{0}}~a\right]r^{\frac{2m}{n-1}}
\end{equation}

 A comparative study of equal powers of $r$ shows that
$m=\frac{n-1}{n-3}$ and $a$ depends on $F_{m}$ and $F_{0}$. Here
for $n=4,m=3$ and this situation is already discussed by Singh et
al [12]. For $n=5$, we have $m=2$ and from (32) we get

\begin{equation}
2b^{2}(4b+\xi)+(2b+\xi)^{2}=0
\end{equation}

where ~~$b=\frac{a}{\sqrt{F_{0}}}$
~~and~~$\xi=\frac{F_{2}}{F_{0}^{2}}$.\\

We note that for real $b$, we must have $b<-\frac{\xi}{4}$ i.e.,
$a<-\frac{F_{2}}{4F_{0}^{3/2}}$, which is essentially the
restriction in (26). It can be shown that the above cubic
equation has at least one positive real root if
$\xi\le-(11+5\sqrt{5})$. Thus, if
$F_{2}\le-(11+5\sqrt{5})F_{0}^{2}$ we have at least one real
positive solution for $a$ which is consistent with equation (29)
(or (32)). Further for $n>5$, we can not have any integral
(positive) solution for $m$ and hence equation (32) is not
consistent for $n>5$. So, it is possible for (at least) radial
null geodesics which initiate from the singularity and reach to
an external observer without get prevented by any trapped surface
for $n\le 5$. Therefore, naked singularity is possible only for
four and five dimensions and for higher dimensions ($n\ge 6$) all
singularities are covered by trapped surfaces leading to black
hole.\\

Further, to examine whether it is possible to have an entire
family of geodesics those have started at the singularity, let us
consider geodesics correct to one order beyond equation (25)
i.e., of the form

\begin{equation}
t=t_{0}+a r^{\alpha}+d r^{\alpha+\beta}
\end{equation}

where as before $a, d, \alpha$ and $\beta$ are positive
constants. Thus equation (29) is modified to

\begin{equation}
a \alpha r^{\alpha-1}+(\alpha+\beta)d
r^{\alpha+\beta-1}=\frac{\left[1-\frac{n-1}{2}\sqrt{F_{0}}~
(t_{0}+a r^{\alpha}+d
r^{\alpha+\beta})-\frac{(2m+n-1)F_{m}}{4\sqrt{F_{0}}}r^{m}(t_{0}+a
r^{\alpha}+d r^{\alpha+\beta})
\right]}{\left[1-\frac{n-1}{2}\sqrt{F_{0}}\left(1+\frac{F_{m}}{2F_{0}}r^{m}
\right)(t_{0}+a r^{\alpha}+d r^{\alpha+\beta})
\right]^{\frac{n-3}{n-1}}}
\end{equation}

So for $\alpha>m$ (retaining terms upto second order) we have

\begin{equation}
\alpha a r^{\alpha-1}+(\alpha+\beta)d
r^{\alpha+\beta-1}=\left(1+\frac{2m}{n-1}\right)\left(-\frac{F_{m}}{2F_{0}}\right)
^{\frac{2}{n-1}}r^{\frac{2m}{n-1}}+D r^{\alpha-\frac{(n-3)m}{n-1}}
\end{equation}

As before, we have the values of $a$ and $\alpha$ in equation
(31) and\\

$$\beta=1-\frac{(n-3)m}{n-1}~~~\text{and} ~~~
d=\frac{D}{2+\frac{(5-n)m}{n-1}}$$ ~~with

\begin{equation}
D=\frac{1}{2}\sqrt{F_{0}}\left(-\frac{F_{m}}{2F_{0}}\right)
^{\frac{5-n}{n-1}}\left\{(n-3)\left(1+\frac{2m}{n-1}\right)-n+1\right\}.
\end{equation}

As we have similar conclusion as before for $\alpha>m$ so we now
consider the case $\alpha=m$. But if we restrict ourselves to the
five dimensional case then $m=\alpha=2$ and we have the same cubic
equation (33) for $b$. Now $\beta$ can be evaluated from the
equation

\begin{equation}
2+\beta=\frac{2^{5/2}b}{(-\xi-4b)^{\frac{3}{2}}}
\end{equation}

and we must have $\beta>0$, otherwise the geodesics will not lie
in the real space-time. As there is no restriction on $d$ so it
is totally arbitrary. This implies that there exists an entire
family of outgoing null geodesics terminated in the past at the
singularity for four and five
dimensions only.\\

\section{\normalsize\bf{Discussion and Concluding Remarks}}

In this paper, we have studied spherical dust collapse in an
arbitrary $n$ dimensional space-time. We have considered both
non-marginal and marginally bound cases separately in sections II
and III respectively. For non-marginal case we have seen that
naked singularity  may be possible for all dimensions ($n\ge 4$).
However, to get a definite conclusion about naked singularity we
should study the geodesic equations as it has been done in section
III for marginally bound case. But we can not proceed further due
to the presence of the complicated hypergeometric functions.
Therefore, no definite conclusion is possible for non-marginal
case.\\

On the other hand, for marginally bound case we have definitely
concluded that naked singularity is possible only for $n\le 5$ by
studying the existence of radial null geodesic through the
singularity. This result also supports our earlier speculation
(see ref.[14]). Here, we should mention that this result depends
sensitively on the choice of the initial conditions. In fact, if
we do not assume the initial density to have an extremum value at
the centre (i.e., $\rho_{1}\ne 0$) the naked singularity will be
possible in all dimensions.\\

Finally, we should mention that the naked singularity described
above is only a local feature, it is not at all a global aspect
i.e., it violates the strong form of CCC. For future work it will
be nice to consider the non-marginal case more extensively so
that some definite conclusion can be drawn regarding the state of
the singularity. Also it will be interesting to study
gravitational collapse for perfect fluid model.\\\\

{\bf Acknowledgement:}\\

The authors are thankful to the members of Relativity and
Cosmology Research Centre, Department of Physics, Jadavpur
University for helpful discussion. Part of this work has been
done during a visit to IUCAA, Pune. One of the authors (U.D) is
thankful to CSIR (Govt. of India) for awarding a Junior Research
Fellowship.\\

{\bf References:}\\
\\
$[1]$  S.W. Hawking and G.F.R. Ellis, The large scale structure
of space-time (Cambridge. Univ. Press,
Cambridge, England, 1973).\\
$[2]$  R. Penrose, {\it Riv. Nuovo Cimento} {\bf 1} 252 (1969);
in General Relativity, an Einstein
Centenary Volume, edited by S.W. Hawking and W. Israel (Camb. Univ. Press, Cambridge, 1979).\\
$[3]$  For recent reviews, see, e.g. P.S. Joshi, {\it Pramana}
{\bf 55} 529 (2000); C. Gundlach, {\it Living Rev. Rel.} {\bf 2}
4 (1999); A. Krolak, {\it Prog. Theo. Phys. Suppl.} {\bf 136} 45
(1999); R. Penrose, in Black holes and relativistic stars, ed. R.
M. Wald (Univ. of Chicago Press, 1998);
T.P.Singh, {\it gr-qc}/9805066.\\
$[4]$  P.S. Joshi, Global Aspects in Gravitation and Cosmology (Oxford Univ. Press, Oxford, 1993).\\
$[5]$ C.J.S. Clarke, {\it Class. Quantum Grav.} {\bf 10} 1375
(1993);T.P. Singh, {\it J. Astrophys. Astron.}
{\bf 20} 221 (1999).\\
$[6]$  F.C. Mena, R. Tavakol and P.S. Joshi, {\it Phys. Rev. D} {\bf 62} 044001 (2000).\\
$[7]$  P.S. Joshi, N. Dadhich and R. Maartens, {\it Phys. Rev. D} {\bf 65} 101501({\it R})(2002).\\
$[8]$  A. Benerjee, A. Sil and S. Chatterjee, {\it Astrophys. J.}
{\bf 422} 681 (1994); A. Sil and S. Chatterjee, {\it Gen. Rel.
Grav.} {\bf 26} 999 (1994); S. Chatterjee, A. Banerjee and B.
Bhui, {\it Phys. Lett. A} {\bf 149} 91 (1990).\\
$[9]$  S. G. Ghosh and N. Dadhich, {\it Phys. Rev. D} {\bf 64}
047501 (2001).\\
$[10]$  S. G. Ghosh and A. Beesham, {\it Phys. Rev. D} {\bf 64}
124005 (2001); {\it Class. Quantum Grav.}
{\bf 17} 4959 (2000).\\
$[11]$  S. G. Ghosh and A. Banerjee, {\it Int. J. Mod. Phys.
D},(2002) (accepted),
{\it gr-qc}/0212067 (2002).\\
$[12]$  S. Barve, T. P. Singh, C. Vaz and L. Witten, {\it Class.
Quantum Grav.} {\bf 16} 1727 (1999).\\
$[13]$ T. Harada, H. Iguchi and K.I. Nakao, {\it Prog. Theor.
Phys.} {\bf 107} 449 (2002) .\\
$[14]$ A. Banerjee, U. Debnath and S. Chakraborty, {\it gr-qc}/0211099 (2002) .\\

\end{document}